\documentclass[aps,prl,twocolumn,showpacs,10pt]{revtex4-1}
\usepackage{amssymb}
\usepackage{amsmath}
\usepackage[pdftex]{graphicx}
\usepackage{dsfont}
\usepackage{bm}
\usepackage[normalem]{ulem}
\usepackage{hyperref}
\hypersetup{
    colorlinks,%
    citecolor=blue,%
    filecolor=blue,%
    linkcolor=blue,%
    urlcolor=blue
}

\providecommand{\brho}{\bm{\rho}} 
\providecommand{\ket}[1]{\vert #1\rangle} 
\providecommand{\bra}[1]{\langle #1\vert} 
\providecommand{\mean}[1]{\langle #1 \rangle} 

\usepackage{color}

\begin{document}
\title{Magnetically Defined Qubits on 3D Topological Insulators}
\author{Gerson J.~Ferreira}
\author{Daniel Loss}
\affiliation{Department of Physics, University of Basel, Klingelbergstrasse 82, CH-4056 Basel, Switzerland}
\date{\today}

\begin{abstract}
We explore potentials that break time-reversal symmetry
to confine the surface states of 3D topological insulators into quantum wires and quantum dots. A magnetic domain wall on a ferromagnet insulator cap layer provides interfacial states predicted to show the quantum anomalous Hall effect (QAHE). Here we show that confinement can also occur at magnetic domain heterostructures, with states extended in the inner domain, as well as interfacial QAHE states at the surrounding domain walls. The proposed geometry allows the isolation of the wire and dot from spurious circumventing surface states. For the quantum dots we find that highly spin-polarized quantized QAHE states at the dot edge constitute a promising candidate for quantum computing qubits.
\end{abstract}
\maketitle

\paragraph{Introduction.} A 3D topological insulator (TI) is characterized by a gapped bulk band structure, and a gapless dispersion of surface states, with low-energy excitations described by the Dirac equation \cite{Kane2010Review,ShouCheng2008,Ando2013Review,Volkov1,Volkov2,VokovInBook,Fu2007TI3D,Nitin2012TIevidence}.
The strong spin-orbit interaction (SOI), responsible for such exotic surface states, makes TIs interesting for spintronics applications \cite{Molenkamp2007QSHE,Yaroslav2012ThinFilm,Paudel2012TIQDot,Paudel2012Optical,Fuhrer2011TIFET,Fuhrer2012TIDot}.
For this purpose it is desirable to introduce and control a gap into the Dirac cone, which requires potentials that break time-reversal symmetry (TRS) \cite{Kane2010Review,ShouCheng2008,Pankratov1987QAHE,Berry1987Neutrino,Alonso1997Boundary,McCann2004Boundary,Nitin2012Hedgehog,Nitin2012TIGap}.
In graphene, magnetic confinement can be obtained by engineering a nonuniform vector potential \cite{DeMartinoPRL2007MagnConfGraphene}. 
In TIs, one possible mechanism is the exchange coupling induced by a ferromagnet insulator (FMI) deposited on top of a TI \cite{Nitin2012Hedgehog,Nitin2012TIGap,Timm2012TIQDot}. In the quantum anomalous Hall effect (QAHE) \cite{Pankratov1987QAHE,Yu2010QAHE} TI states confined along a domain wall of the FMI are helical, and carry a dissipationless current. 
These correspond to one-half of the quantum spin Hall effect \cite{Molenkamp2012Imaging,Molenkamp2013Spatially}.
Experimental observation of the QAHE was recently discussed in Ref.~\cite{Chang2013QAHE}. 

\begin{figure}[t]
 \centering
 \includegraphics[width=\columnwidth,keepaspectratio=true]{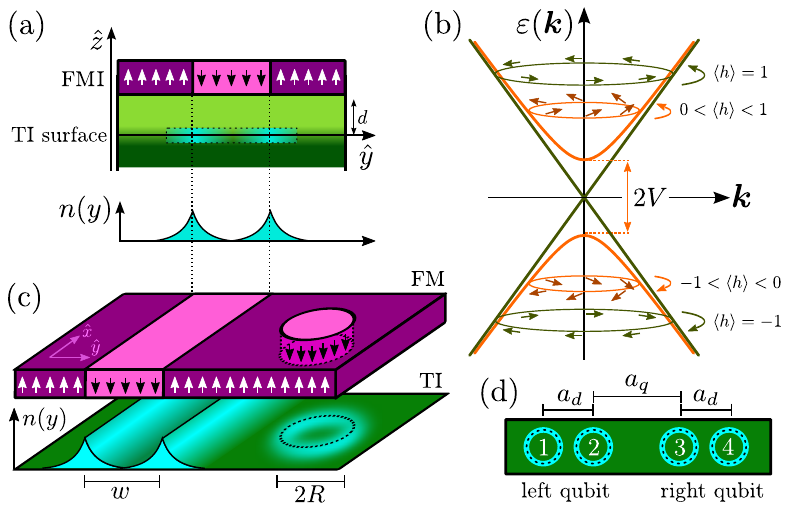}
 \caption{(a) Surface states of 3D TIs confined by domains of a FMI  that  create a TRS-breaking potential $V$ through exchange coupling.
 The bare TI spectrum $V=0$ is a helical Dirac cone with a Rashba spin orientation $\hat{k}\times\hat{z}$. The  potential $V \neq 0$ tilts the spins out of the plane. The mean value of the helicity operator $\mean{h}$ quantifies the deviation from the helical case. (c) Quantum wire and dot defined by the magnetic heterostructure pattern. (d) Arrangement of four dots defining two qubits.}
 \label{fig:system}
\end{figure}

In this work, we explore gapped 3D TI surface states to define quantum wires and quantum dots beyond the domain-wall-induced interfacial states. For concreteness, we consider the exchange coupling induced by a FMI cap layer \cite{Nitin2012Hedgehog,Nitin2012TIGap} as the TRS-breaking potential, see Fig.~\ref{fig:system}. We show that the confinement of the surface states follows the magnetization domains' pattern (magnetic heterostructures), with the interfacial QAHE states as a particularly interesting case. This geometry protects the system against spurious circumventing surface states.
We show that the interfacial QAHE states are highly spin polarized due to a constraint imposed by the hard-wall boundary conditions \cite{Berry1987Neutrino,Alonso1997Boundary,McCann2004Boundary}, and a generalization to realistic soft-wall potentials only slightly relaxes this constraint.
For quantum dots, we find quantized interfacial QAHE states, which constitute promising candidates for quantum computing qubits. The high spin polarization of these states, and the pure magnetic confinement potentially suppress effects from nonmagnetic perturbations.

\paragraph{Hamiltonian \& Helicity.} We consider the 2D Dirac Hamiltonian for the surface states of a 3D TI \cite{Fu2007TI3D,Moore2007TI3D,Kane2010Review}

\begin{equation}
 H = v_F \bm{\sigma}\cdot\bm{\pi} + V(\brho) + \gamma_z B \sigma_z,
\end{equation}
with Fermi velocity $v_F$, Pauli matrices $\bm{\sigma} = (\sigma_x, \sigma_y, \sigma_z)$, conjugate momentum $\bm{\pi} = \bm{p} + e\bm{A}$, $\bm{p} = (p_x, p_y)$, and $\brho = (x, y)$. The spin operator is $\bm{S} = (\hbar/2)(\sigma_y, -\sigma_x, \sigma_z)$.
For the Fock-Darwin states discussion, we choose the symmetric gauge $\bm{A} = B \rho \hat{\theta}/2$ in polar coordinates ($\bm{B} = \nabla \times \bm{A} = B\hat{z}$). The external potential $V(\brho)$ is a $2\times 2$ matrix and will be discussed later on. The last term is the Zeeman splitting with gyromagnetic ratio $\gamma_z$.

The Dirac spectrum of $H$ with $V(\brho) = 0$ and $\gamma_z = 0$ is helical; see Fig.~\ref{fig:system}(b). The helicity operator $h = 2(\bm{S}\times \hat{p})\cdot\hat{z} = 2S_t$ measures the in-plane spin projection transversal to the momentum. For helical states, $[H, h]=0$. The eigenstates shown in Fig.~\ref{fig:system}(b) have energies $\varepsilon_\pm(\bm{k}) = \pm \sqrt{(\hbar v_F k)^2 + (\gamma_z B)^2}$ [for $V(\brho)=0$, $\bm{A} = 0$]. The corresponding eigenstates are

\begin{equation}
 \psi_\pm(\brho) =
\begin{pmatrix}
 \hbar v_F k_- \\
 \varepsilon_\pm - \gamma_z B
\end{pmatrix}
e^{i\bm{k} \cdot \brho},
\label{eq:planewaves}
\end{equation}
with $k_- = k_x -i k_y$. For $\gamma_z B = 0$ the spins lie in the $xy$ plane with a Rashba orientation. A finite $\gamma_z B \neq 0$ breaks TRS, opening a $2\gamma_z B$ gap. In this case $[H,h]\neq 0$, and $|\mean{h}| < 1$ quantifies the deviation from helical states as the spins tilt out of the plane. This \textit{hedgehog spin texture} was recently observed \cite{Nitin2012Hedgehog,Nitin2012TIGap}.

\paragraph{Hard- \& Soft-Wall Potentials.} The hard-wall boundary conditions for the Dirac equation were extensively discussed in the literature \cite{Berry1987Neutrino,Alonso1997Boundary,McCann2004Boundary}. Because of the first-order derivatives in the kinetic operator, the spinor is discontinuous across a hard wall \cite{Alonso1997Boundary}. McCann and Fal'ko \cite{McCann2004Boundary} established a classification of matrices for hard-wall confinement. Here, we follow a slightly different derivation that allows an immediate generalization to define soft-wall confining matrix potentials $V(\brho)$.

One can consider $H$ with $\bm{A} = 0$ and $\gamma_z B = 0$ without loss of generality. We write the general potential as

\begin{equation}
 V(\brho) = V_0 \tilde{M} \Theta(\rho - \rho_B),
\end{equation}
where $V_0$ is the scalar intensity, $\tilde{M}$ is a unitary Hermitian matrix, and $\Theta(\rho - \rho_B)$ is the step function defining the boundary at $\rho=\rho_B$ with the coordinates along the normal unit vector $\hat{n}_B$. In the hard-wall limit, $V_0 \rightarrow \infty$, the spinor discontinuity at the interface reads $\psi(\brho) \approx \psi(\brho_B) [1-\Theta(\rho - \rho_B)]$. Consequently, $\nabla \psi(\brho) \approx -\psi(\brho_B) \delta(\rho-\rho_B)\hat{n}_B$, and $V(\brho)\psi(\brho) \approx \hbar v_F \tilde{M}\psi(\brho_B)\delta(\rho-\rho_B)$. Integrating $H$ along $\hat{n}_B$ across the boundary, we obtain the hard-wall boundary conditions \cite{McCann2004Boundary}

\begin{equation}
 \left(\mathds{1} - i\sigma_B\tilde{M}\right)\psi(\brho_B) = 0,
 \label{eq:hardwallBC}
\end{equation}
with $\sigma_B = \hat{n}_B\cdot\bm{\sigma}$. Equation \eqref{eq:hardwallBC} admits nontrivial solutions $\psi(\brho_B) \neq 0$ only if $(\mathds{1} - i\sigma_B\tilde{M})$ is singular, which requires $\{\tilde{M}, \sigma_B\} = 0$ and $\tilde{M}^2 = \mathds{1}$. Soft-wall potentials (finite $V_0$) defined by matrices $\tilde{M}$ that satisfy the above conditions show confined spinors, continuous at the interface $\brho = \brho_B$, and with penetration length $\ell = \hbar v_F/V_0$. The discontinuity is recovered as $\ell \rightarrow 0$ in the hard-wall limit.

For a quantum wire along $\hat{x}$, $\sigma_B = \pm \sigma_y$ and the above conditions give $\tilde{M} = \sigma_z$ or $\sigma_x$. For a circular quantum dot, $\sigma_B = \sigma_r$ (radial), the requirement is $\tilde{M} = \sigma_z$ or $\sigma_\theta$ (polar). The cases $\sigma_x$ and $\sigma_\theta$ correspond to the Landau level terms from $\bm{A} = -y B \hat{x}$ (wire), and $\bm{A} = \frac{1}{2}B\rho\hat{\theta}$ (dot), both yielding $\bm{B} = B\hat{z}$. The $\tilde{M} = \sigma_z$ potentials can be implemented by a nonuniform Zeeman term or a local exchange coupling with a FMI cap layer, as in Fig.~\ref{fig:system}.

\paragraph{Quantum wire.} For simplicity consider $H$ with $\bm{A} = 0$ and $\gamma_z B = 0$. The soft-wall confinement potential is

\begin{equation}
  V(y) = \left\{
\begin{array}{r l}
 V_i\sigma_z & \text{ for } |y| < w/2\\
 V_o\sigma_z & \text{ for } |y| \geq w/2,
\end{array}
\right.
\label{eq:wireV}
\end{equation}
where $V_i$ and $V_o$ are the amplitudes inside and outside the wire of width $w$. The solutions of each piecewise region (labeled by $j$) are given by Eq.~\eqref{eq:planewaves}, replacing $k_y \rightarrow k_j = \pm \sqrt{(\varepsilon^2-V_j^2)/(\hbar v_F)^2 - k_x^2}$. The local band structure of each region is equivalent to Fig.~\ref{fig:system}(b) with a $2V_j$ gap. The wire band structure is obtained by imposing the spinor continuity at the interfaces $y = \pm w/2$, with evanescent solutions on outer regions ($|\varepsilon| < |V_j|$ for $|y| \geq w/2$).

\begin{figure}[t]
 \centering
 \includegraphics[width=\columnwidth,keepaspectratio=true]{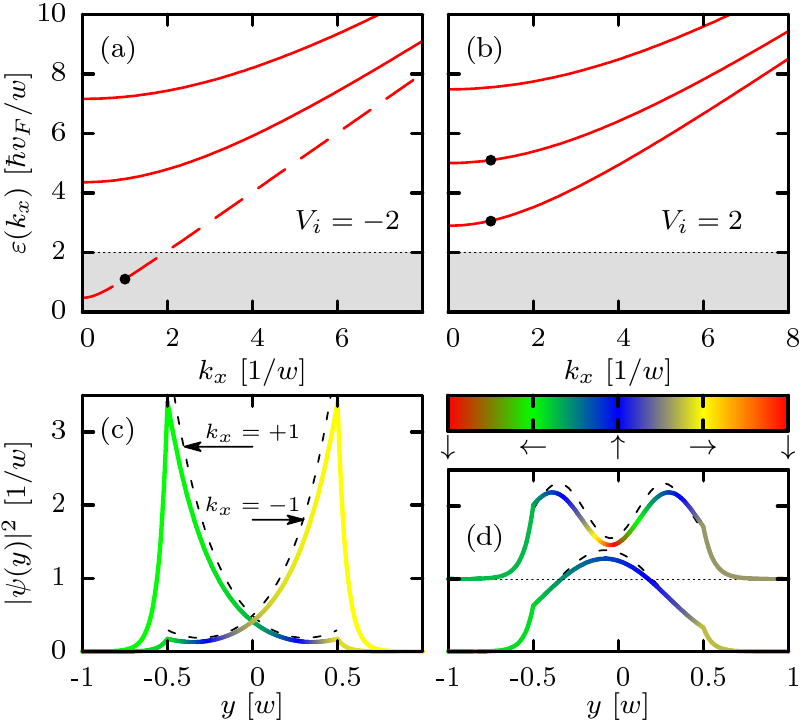}
 \caption{(a),(b) Band structure of the quantum wire for $V_o = 10$~$\hbar v_F/w$ and $V_i = \pm 2$~$\hbar v_F/w$. The gray area delimits the gap of the inner region. (a) Potentials $V_i$ and $V_o$ with opposite signs lead to a band (dashed line) of interfacial states at the edges. (c)-(d) Density $|\psi(y)|^2$ for the $k_x= 1/w$ states of the lowest bands [indicated by black dots in (a) and (b)]. The density of the second band is shifted upwards for clarity. The color code represents the spin texture along the $S_y$--$S_z$ plane [($\leftarrow$, $\rightarrow$), and ($\uparrow$, $\downarrow$), respectively], locked by SOI in the Dirac equation.
 The dashed lines show the hard-wall solutions for comparison.}
 \label{fig:wire}
\end{figure}

Figure \ref{fig:wire} shows the wire energy dispersion for $V_o = 10$~$\hbar v_F/w$, and $V_i$ indicated on the panels. The sign change between $V_i$ and $V_o$ in Fig.~\ref{fig:wire}(b) is equivalent to the band inversion in TI and leads to localized interfacial states [Fig.~\ref{fig:wire}(c)] within the gap region (gray area). These are the QAHE states \cite{Pankratov1987QAHE,Yu2010QAHE,Kane2010Review,Molenkamp2012Imaging,Molenkamp2013Spatially,Chang2013QAHE}. The other branches correspond to normal, nontopological, states localized within the full inner domain region, as shown in Fig.~\ref{fig:wire}(d).

Since the momentum is along $\hat{x}$, the SOI locks the spin into the $S_y$--$S_z$ plane. Here, we use ($\leftarrow$, $\rightarrow$) to refer to the projections along $S_y$, and ($\uparrow$, $\downarrow$) for $S_z$. In the hard-wall limit, the boundary condition given by Eq.~\eqref{eq:hardwallBC} implies that at $y=+(-) w$, the local spin is $\leftarrow$ ($\rightarrow$). The QAHE states are localized at these edges, and in the strong confinement limit (large $V_i$ or wide wire), their spin projection approaches full in-plane polarization ($\leftarrow$ or $\rightarrow$), thus reaching the helical regime $|\mean{h}| = 1$ and suppressing the gap in Fig.~\ref{fig:wire}(a). The softwall slightly relaxes this condition, but still shows 
such a spin constraint; see Figs.~\ref{fig:wire}(c) and \ref{fig:wire}(d). More generally, the spatial spin texture follows the color-code diagram, and the number of rotations between the edges increases with the band index; see Fig.~\ref{fig:wire}(d).

\paragraph{Quantum dots.} To solve $H$ for a quantum dot in polar coordinates $x=\rho\cos\theta$ and $y=\rho\sin\theta$, the kinetic term has to be symmetrized

\begin{equation}
 v_F\bm{\sigma}\cdot\bm{\pi} \rightarrow v_F\big(\sigma_r p_r + \sigma_\theta p_\theta\big) + i\hbar v_F \frac{\sigma_r}{2r} + \hbar\omega_B \frac{\rho\sigma_\theta}{2\ell_B},
 \label{eq:symmetrized}
\end{equation}
where $p_r$ and $p_\theta$ are components of the momentum operator, $\omega_B = v_F/\ell_B$ is the cyclotron frequency, and $\ell_B = \sqrt{\hbar/eB}$ is the magnetic length. The radial and polar Pauli matrices are  $\sigma_r = \sigma_x\cos\theta + \sigma_y\sin\theta$ and $\sigma_\theta = -\sigma_x\sin\theta + \sigma_y\cos\theta$. In Eq.~\eqref{eq:symmetrized}, the second term arises from the symmetrization, and the last term from the symmetric gauge, responsible for the Landau levels (LLs).
The dot radial soft-wall potential $V(\rho)$ has the same form of Eq.~\eqref{eq:wireV}, with the inner and outer regions delimited by the radius $R$.

The z component of the total angular momentum ($J_z = L_z + S_z$, and $L_z = -i\hbar\partial_\theta$) commutes with $H$. The common set of eigenstates yields $\psi_m(\rho,\theta) = \varphi_m(\theta) \psi_{m}(\rho)$, with a diagonal matrix $\varphi_m(\theta) = \text{diag}[e^{i m \theta}, e^{i(m+1)\theta}]$. The integer $m$ defines the eigenvalues $(m+\frac{1}{2})\hbar$ of $J_z$. For $B=0$, the radial solutions are

\begin{equation}
  \psi^j_{m}(\rho) =
  \begin{bmatrix}
 \sqrt{\rho} Q^j_m(k_j\rho) \\
 \dfrac{i\hbar v_\perp k_j}{\varepsilon+V_j} \sqrt{\rho} Q^j_{m+1}(k_j\rho)
\end{bmatrix},
\end{equation}
where $j$ labels the inner and outer regions, $\hbar v_F k_j = \sqrt{\epsilon^2 - V_j^2}$ and $Q^i_m(x) = J_m(x)$ and $Q^o_m(x) = H^{(1)}_m(x)$ are the Bessel and the Hankel functions of the first kind. For finite $B$, the solutions are given by Kummer $M$ and $U$ functions (not shown). The eigenstates are found imposing continuity at the interface $\rho = R$.

\begin{figure}[t]
 \centering
 \includegraphics[width=\columnwidth,keepaspectratio=true]{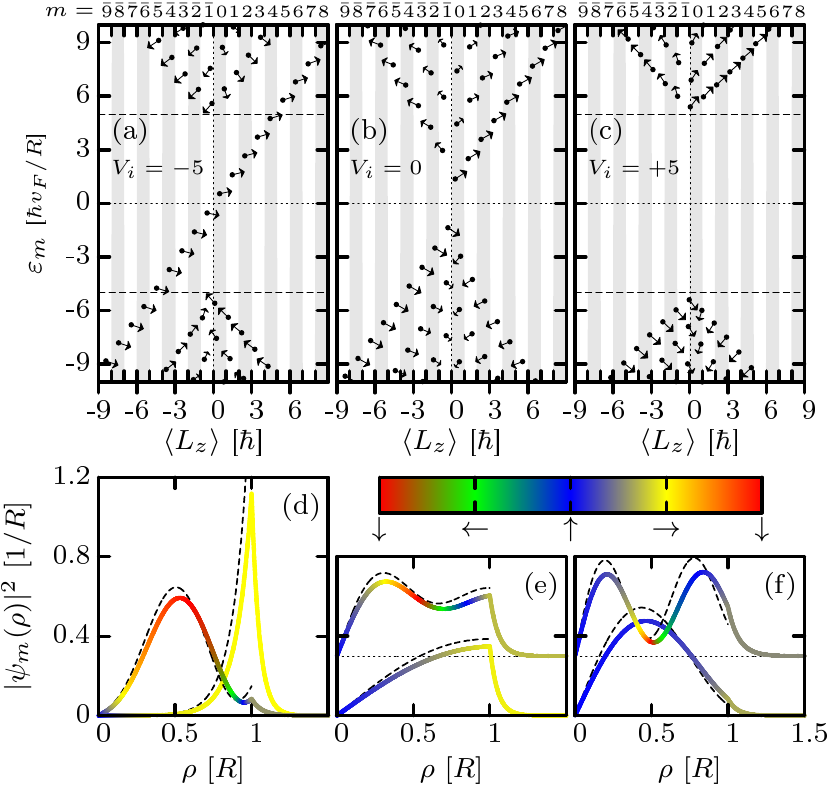}
 \caption{(a), (c) Eigenenergies as a function of $\mean{L_z}$ for $V_o = 10$ and $V_i$~$[\hbar v_\perp/R]$ indicated in the panels. The gray or white stripes delimit different values $m$ (on top). The arrows represent the average spin projections $\mean{S_\nu}$ (log scale) in the $S_r$--$S_z$ plane [($\leftarrow$, $\rightarrow$) and ($\uparrow$, $\downarrow$), respectively]. (a) For $V_i V_o < 0$, the diagonal branch of interfacial states corresponds to a quantization of the QAHE wire states.
 \mbox{(d), (f)} Corresponding densities $|\psi_m(\rho)|^2$ for $m=0$ of the first and second positive energy states in \mbox{(a)-(c) } (some shifted for clarity). The dashed lines are the hard-wall solutions for comparison.}
 \label{fig:dot}
\end{figure}

The eigenenergies are shown in Figs.~\ref{fig:dot}(a)-\ref{fig:dot}(c) as a function of $\mean{L_z}$ or $m$. For $V_i V_o < 0$, a branch of interfacial states is present [Figs.~\ref{fig:dot}(a) and \ref{fig:dot}(d)], corresponding to a quantization of the QAHE wire states from the domain wall at the dot edge. The SOI constrains the spatial spin texture to be along the $S_r$--$S_z$ plane. At $\rho = 0$, the spin can only be $\uparrow$ or $\downarrow$ due to symmetry, and at $\rho = R$, the hard-wall boundary condition imposes spin $\rightarrow$ or $\leftarrow$.
As for the wire, in the strong confinement limit, the interfacial states approach the helical regime ($|\mean{h}| = 1$) as the spin becomes fully in plane. These are the states we argue to be promising qubit candidates.

\begin{figure}[t]
 \centering
 \includegraphics[width=\columnwidth,keepaspectratio=true]{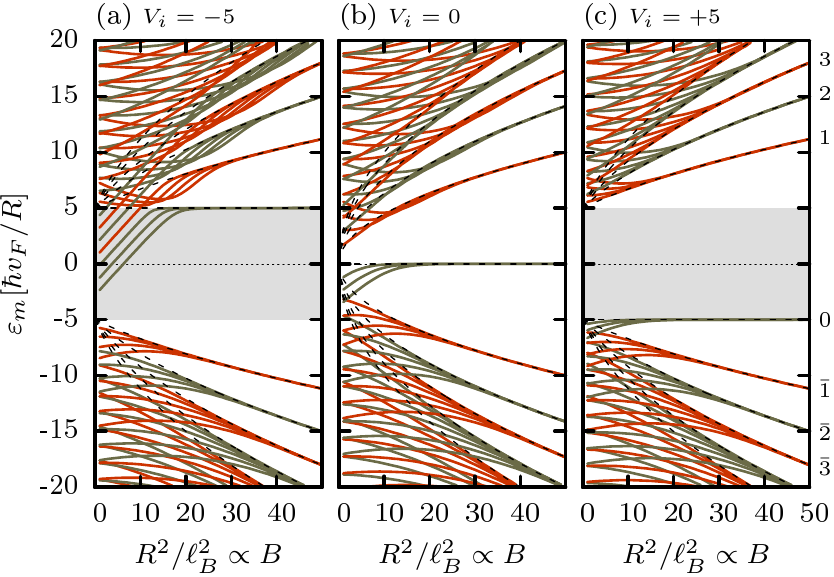}
 \caption{TI Fock-Darwin states ($-3 \leq m \leq 3$) converging into Landau levels as a function of the magnetic field parametrized by $R^2/\ell_B^2$. Here, we consider hard-wall boundary conditions $V_o \rightarrow \infty$ and $V_i$ as indicated in each panel.
 (a)  With increasing $B$, the interfacial states are expelled from the gap region (gray area), as the LL confining potential becomes dominant over $V(\rho)$.
 For high magnetic fields, all cases converge to the LL spectrum $\varepsilon_{n_{LL}} = \pm \hbar\omega_B\sqrt{2 n_{LL} + (V_i/\hbar\omega_B)^2}$ for $n_{LL} \neq 0$, and $\varepsilon_0 = -V_i$ for $n_{LL} = 0$ (dashed lines).
 }
 \label{fig:fockdarwin}
\end{figure}

\paragraph{TI Fock-Darwin \& Landau level states.} As in the normal Fock-Darwin states, at low $B$, the quantum dot confinement $V(\rho)$ is dominant, while at high $B$, the vector potential term leads to the highly degenerate LLs; see Fig.~\ref{fig:fockdarwin} (for $\gamma_z = 0$).
The LL confinement is \textit{normal}; i.e., it does not contain a gap inversion. Therefore, the interfacial states present for $V_i V_o < 0$ at low $B$ are expelled from the gapped region (gray area) as the LL confinement becomes dominant.

\paragraph{Two-qubit gates.} Consider the linear arrangement of four quantum dots in Fig.~\ref{fig:system}(d), where each qubit is defined by a pair of dots with states from the interfacial QAHE branch; see Figs.~\ref{fig:dot}(a) and \ref{fig:dot}(d). Their energy separation defines a temperature energy scale $k_B T \ll \hbar v_F/R$, which avoids coupling to other states in this branch.

Each pair of dots containing a single electron within this subspace is described by a $2\times 2$ effective Hamiltonian

\begin{equation}
 H_\text{qubit} = \Delta_d \tau_x + \delta_d \tau_z,
\end{equation}
where $\bm{\tau} = (\tau_x, \tau_y, \tau_z)$ are Pauli matrices acting on single-particle states localized on each dot. The single-particle hybridization energy $\Delta_d$ and the dot-energy detuning $\delta_d = (\varepsilon_1-\varepsilon_2)/2$ [or $=(\varepsilon_3-\varepsilon_4)/2$] can be controlled by electrostatic gates and electric fields, respectively.

To derive an effective qubit Hamiltonian $H_Q$ for two particles in four dots, we label the basis of Slater determinants, with the particles at sites $i$ and $j$, as $\ket{s_{ij}}$. Since the single-particle spinors are highly localized, and considering the interdot distances $a_d < a_q \ll \ell$, the Coulomb interaction reduces to a simple on-site repulsion description; thus, it is diagonal in the localized $\ket{s_{ij}}$ basis. The diagonal matrix elements reads $\mathcal{D}_{ij} = \bra{s_{ij}} U_C \ket{s_{ij}} \approx e^2/\kappa r_{ij}$, where $\kappa$ is the dielectric constant and $r_{ij}$ the distance between dots $i$ and $j$. Moreover, we consider a regime where all $\mathcal{D}_{ij}$ dominate over the single-particle hybridization energy $\Delta_d$. The condition $a_d < a_q$ leads to high charging energies per qubit, $\mathcal{D}_{12} = \mathcal{D}_{34} \gg \text{other } \mathcal{D}_{ij}$, allowing us to neglect the doubly occupied states. Within the reduced basis $\{\ket{s_{14}}, \ket{s_{13}}, \ket{s_{24}}, \ket{s_{23}}\}$, we obtain

\begin{equation}
 H_Q =
\begin{pmatrix}
E_{00} & 0 & 0 & 0\\
0 & E_{01} & \Delta & 0\\
0 & \Delta & -E_{01} & 0\\
0 & 0 & 0 & E_{11}
\end{pmatrix}.
\end{equation}

The matrix elements in $H_Q$ are

\begin{eqnarray}
 E_{00} &=& -C_{00}-\Delta +\delta_a/2,\\
 E_{01} &=& \delta_b/2,\\
 E_{11} &=& C_{11}-\Delta - \delta_a/2,\\
 \Delta &=& \left[\frac{2 \Delta_d^2}{2 C_{00}-\delta_a}-\frac{2 \Delta_d^2}{2 C_{11}-\delta_a}\right],
\end{eqnarray}
where $C_{00} = \mathcal{D}_{13}-\mathcal{D}_{14}$ and $C_{11} = \mathcal{D}_{23}-\mathcal{D}_{13}$. Weak electric fields applied at each dot can control the independent parameters $\delta_a$ and $\delta_b$ defined by the dot detuning $\delta_a = \varepsilon_1 - \varepsilon_3 + \varepsilon_4 - \varepsilon_2$ and $\delta_b = \varepsilon_1 + \varepsilon_3 - \varepsilon_4 - \varepsilon_2$. The central block of $H_Q$ has eigenenergies $\pm \hbar\omega_{01} = \pm \sqrt{\delta_b^2/4 + \Delta^2}$.

Because of the strong Coulomb repulsion, the ground state of the system is $\ket{s_{14}}$, where the particles are repelled to the outer dots. The higher-energy state is $\ket{s_{23}}$ with the particles in the inner dots. The other two states have similar energies due to the symmetry ($\ket{s_{13}}$ and $\ket{s_{24}}$ are mirrored) and hybridize. This motivates the choice of logical ``0'' and ``1'' qubit states as $\ket{00} = \ket{s_{14}}$, $\ket{01} = \ket{s_{13}}$, $\ket{10} = \ket{s_{24}}$, and $\ket{11} = \ket{s_{23}}$.

Assuming a rectangular pulse control of the interaction parameters, the time evolution takes the form $U(\tau) = \exp[-i H_Q \tau /\hbar]$. This defines a controlled phase-flip (CPF) gate $U(\tau) = \text{diag}[1, 1, 1, -1]$, for an operation time $\tau = 2 \pi n_1/\omega_{01}$, if the detuning parameters $\delta_a$ and $\delta_b$ are set to satisfy

\begin{equation}
 \frac{E_{00}}{\omega_{01}} = \frac{-n_2}{n_1}, \; \text{ and } \; \frac{E_{11}}{\omega_{01}} = \frac{n_3 + 1/2}{n_1},
\end{equation}
with integers $n_1$ and $n_2 >0$, and $n_3 \geq 0$.

\paragraph{Energy scales.} The single-particle energy scales are set by $\hbar v_f \approx 300$--$500$~meV~nm for typical materials (Bi$_2$Se$_3$, Pb$_x$Sn$_{1-x}$Te); thus, for wire width $w$ or dot radius $R$ about $100$~nm, the energy scale for the confinement potential lies on the meV range. The on-site Coulomb repulsion is $e^2/\kappa r \approx 1400/\kappa r$~meV for $r$ in nm. Since the dot distance is $r > 2 R$, the energy scales satisfy $e^2/\kappa r \lesssim \hbar v_f/R$ for the dielectric constant $\kappa \gtrsim 2$.

\paragraph{Conclusion.} We considered the confinement of 3D TI surface states by time-reversal-breaking potentials, relaxing the established hard-wall boundary conditions \cite{Berry1987Neutrino,Alonso1997Boundary,McCann2004Boundary} into soft-wall potentials. These can be implemented via local exchange coupling with a ferromagnet insulator cap layer \cite{Yu2010QAHE,Kane2010Review,ShouCheng2008,Nitin2012Hedgehog,Nitin2012TIGap}; see Fig.~\ref{fig:system}. In the proposed heterostructure geometry, the confinement is patterned by magnetic domains built within a larger domain with different magnetization, such that it isolates the system of interest from spurious TI surface states. This is equivalent to the action of split gates on a normal 2D electron gas. 
We expect that the QAHE interfacial states at the edge of quantum dots can potentially be promising candidates for a qubit, since the high spin polarization of the helical regime can potentially suppress the effects of nonmagnetic perturbations. Moreover, the fully magnetic confinement induced by the ferromagnetic domains is less sensitive to electrostatic fluctuations than the usual split-gate electrodes.

\acknowledgments The authors acknowledge support from the Swiss NSF, NCCR Nanoscience, and NCCR QSIT.
\bibliography{tiqubits}

\end{document}